\documentclass[12pt]{iopart} 
\usepackage{iopams}  

\usepackage[english]{babel}
\usepackage{latexsym}
\usepackage{graphics}
\usepackage{subfigure}
\usepackage{epsfig}

\begin{document}

\title[Strongly correlated Fermi-Bose mixtures in disordered optical lattices]{Strongly correlated Fermi-Bose mixtures in disordered optical lattices}

\author{
L. Sanchez-Palencia$^{1,*}$,
V. Ahufinger$^{2}$,
A. Kantian$^{3}$,
J. Zakrzewski$^{4}$,
A. Sanpera$^{5}$, and
M. Lewenstein$^{6,7}$
}

\address{
$^1$Laboratoire Charles Fabry de l'Institut d'Optique,
CNRS and Universit\'e Paris-Sud XI,
B\^at 503, Centre scientifique, F-91403 Orsay cedex, France}
\address{
$^2$ICREA and Grup d'\`Optica, Departament de F\'isica, Universitat Aut\`onoma de Barcelona, E-08193 Belaterra
(Barcelona), Spain}
\address{
$^3$Institut f\"ur Theoretische Physik, Universit\"at Innsbruck, A-6020 Innsbruck, Austria}
\address{
$^4$Instytut Fizyki imienia Mariana Smoluchowskiego i Centrum Bada\'n
Uk\l{}ad\'ow Z\l{}o\.zonych imienia Marka Kaca, Uniwersytet Jagiello\'nski, 
ulica Reymonta 4, PL-30-059 Krak\'ow, Poland}
\address{
$^5$ICREA and Grup de F\'isica Te\`orica, Departament de F\'isica, Universitat Aut\`onoma de Barcelona, E-08193 Belaterra (Barcelona), Spain}
\address{
$^6$ICREA and ICFO-Institut de Ci\`encies Fot\`oniques, Parc Mediterrani de la Tecnolog\'ia, E-08860 Castelldefels (Barcelona), Spain}
\address{
$^7$Institut f\"ur Theoretische Physik, Universit\"at Hannover, D-30167 Hannover, Germany}

\ead{laurent.sanchez-palencia@iota.u-psud.fr}

\date{\today}

\begin{abstract}
We investigate theoretically the low-temperature physics of a two-component
ultracold mixture of bosons and fermions in disordered optical lattices.
We focus on the strongly correlated regime. 
We show that, under specific conditions, composite fermions, 
made of one fermion plus one bosonic hole, form. The 
composite picture is used to derive an effective Hamiltonian whose 
parameters can be controlled via the boson-boson and the boson-fermion 
interactions,
the tunneling terms and the inhomogeneities. We finally investigate the quantum
phase diagram of the composite fermions and we show that it corresponds to the
formation of Fermi glasses, spin glasses, and quantum percolation regimes.
\end{abstract}


\maketitle

\section{Introduction to disordered quantum systems}
\label{sec:introduction}

\subsection{From condensed matter physics to ultracold atomic gases}
\label{sec:introduction:general}
Quantum disordered systems is a very active research field
in condensed matter physics (CM) initiated by the work by P.W.~Anderson \cite{anderson1958} 
who first pointed 
out that quenched ({\it i.e.} time-independent) disorder can dramatically change the properties of a quantum system
compared to its long-range ordered counterpart.
Hence, disorder plays a central role in modern solid state physics as it
can significantly alter 
electronic normal conductivity \cite{localizationbooks}, 
superconductivity \cite{auerbach1994} and
the magnetic \cite{mezard1987,sachdev1999} properties of dirty alloys.

It is by now clear that studying disordered systems is comparable to opening the {\it Pandora box}.
On the one hand, disorder introduces a list of non-negligible difficulties.
First, averaging over disorder usually turns out to be a complex task that requires
either original methods such as 
the replica trick \cite{mezard1987} and
supersymmetry \cite{efetov1997}
or numerical computations using huge samples or a large number of repetitions
with different configurations. 
Second, the possible existence of a huge
number of excited states with infinitely small excitation energies leads
to complex quantum phases such as
bose glasses \cite{fisher1989} or 
spin glasses \cite{mezard1987}.
Third, the interplay of kinetic energy, particle-particle interactions and
disorder is usually a non-trivial problem \cite{shepelyanski1994} 
which is still challenging.
On the other hand, disorder leads to an extraordinary variety of physical phenomena
such as, for example, 
Anderson localization \cite{anderson1958},  
quantum percolation \cite{aharony1994}, or
quantum frustration \cite{misguich2004}.

Studies of quantum disorder in CM have several limitations.
First, the disorder cannot be controlled as it is fixed by the specific realization of the sample. 
In particular, one cannot switch adiabatically from one configuration to another.
Second, the particles are fermions (electrons) and the inter-particle
interaction corresponds to the Coulomb long-range potential.
Third, theoretical studies rely on toy models and experiments do not
provide control parameters.
Thus, it would be highly desirable to consider new kinds of disordered quantum systems.
As shown in this work and also discussed in other papers 
\cite{damski2003,sanpera2004}, one exciting possibility is that of ultracold 
atomic gases where these
problems can be revisited with unique parameters control and measurements
possibilities.

\subsection{Ultracold atomic gases}
\label{sec:introduction:ultracold}
Following the recent progress in cooling and trapping of neutral atoms
\cite{nobel1997}, dilute atomic Bose-Einstein condensates
(BEC) \cite{nobel2001}, degenerate Fermi gases (DFG) 
\cite{fermions} and
mixtures of both \cite{mixtures} are now currently produced at the laboratory.
In these systems almost all parameters prove to be highly controllable 
\cite{natureinsights}:
(i) Using standard techniques, one can easily control the trapping potential,
the size, the density and the temperature of the atomic gas.
(ii) Due to the strong dilution, the contact interactions are usually small
and can be computed {\it ab initio}. In addition, their sign and strength 
can be controlled using Feshbach resonances \cite{feshbach}.
(iii) Quantum statistics is also a degree of freedom as one can use bosons or
fermions.
(iv) Finally, inhomogeneous trapping potentials can be designed almost at will 
using standard optical techniques. 
In particular, periodic potentials (optical lattices) with no defect 
nor phonons can be designed in a wide variety of geometries \cite{grynberg2000}.
Controlled disordered \cite{horak1998} or 
quasi-disordered \cite{guidoni1990s,quasi2005}
potentials can also be optically produced opening new possibilities. 
Hence, equilibrium and transport properties of interacting Bose-Einstein condensates 
in random potentials have been investigated in recent experiments \cite{clement2005,inguscio2005,schulte2005}.

In recent works \cite{sanpera2004,ahufinger2005}, 
we have shown that Fermi-Bose mixtures in an optical lattice with diagonal 
({\it i.e.} with random on-site energy) 
randomness constitute a case study of quantum disordered systems.
In the strongly correlated regime (strong interactions), the dynamics
reduces to a universal effective Hamiltonian whose parameters can be
controlled through the boson-boson and fermion-boson interactions, or
the depth of the periodic and disordered potentials. 
It results in a rich quantum phase diagram ranging from 
Fermi glasses and Fermi Anderson localization to 
quantum spin glasses and quantum percolation \cite{sanpera2004}.


This paper reviews our works on strongly correlated Fermi-Bose mixtures 
in disordered optical lattices. It further details the results 
discussed in Ref.~\cite{sanpera2004}. In Ref.~\cite{ahufinger2005}, we have 
presented a complete study of the system and discussed a wide variety of cases,
including several composite types and quantum phases. As in such a rich system,
completeness competes with conciseness, we think that a shorter review would
be useful for non-specialist readers.
It is the aim of this paper to provide such a concise review of our findings.
We have thus chosen to restrict the present analysis to a specific case that 
proves to be non the less paradigmatic but also one of the richest ones. 
For more details, the interested reader should
refer to   Ref.~\cite{ahufinger2005}.

The paper is organized as follows. 
In section~\ref{sec:strongly}, we introduce the model describing the 
considered system (\ref{sec:FBHm}) and the composite fermions 
formalism that leads to the effective hamiltonian (\ref{sec:composites} and \ref{sec:effective}). Section~\ref{sec:weak} shows the results for the non-disordered and weakly disordered lattices and 
in section~\ref{sec:spinglass} the strong disorder limit, the spin glass limit, is discussed. Finally, we summarize and
discuss our results in section~\ref{sec:conclusion}

\section{Strongly correlated composite fermions in inhomogeneous optical lattices}
\label{sec:strongly}

\subsection{The Fermi-Bose Hubbard Hamiltonian}
\label{sec:FBHm}
Consider a mixture of ultracold bosons (b) and spinless (or spin-polarized)
fermions (f) trapped in an optical lattice.
In addition, the mixture is subjected to on-site inhomogeneities consisting in a 
harmonic confining potential and/or in diagonal disorder.
Eventually, the periodic potential and the inhomogeneities are different
for the bosons and the fermions. However, we assume that the lattices for the 
two species have the same periodicity. The lattice sites are indexed by $i$
and are thus the same for the fermions and the bosons.
In all cases considered below, the temperature is assumed to be low enough and the potential wells deep enough so that only quantum states in the fundamental Bloch bands for both
the bosons and the fermions are populated. Notice that, this requires that the filling factor for fermions $\rho_\textrm{f}$ is smaller than $1$ {\it i.e.} the total number of fermions $N_\textrm{f}$ is smaller than the total number of lattice sites $N$.

We use the Wannier basis of the fundamental Bloch band which corresponds to 
wave-functions well localized in each lattice site \cite{solids} for
both the bosons and the fermions. 
This basis is particularly well suited for the strongly correlated regime that is
investigated here \cite{jaksch1998}. Then, the second quantization Hamiltonian reduces to the 
Fermi-Bose Hubbard (FBH) model
\cite{auerbach1994,sachdev1999,other}~:
\begin{eqnarray}
H_\textrm{FBH} & = &
-\sum_{\langle ij \rangle}\left[ J_\textrm{b} b^{\dagger}_i b_j+J_\textrm{f} f^{\dagger}_i f_j + h.c. \right] \nonumber \\
&& + \sum_i \left[\frac{V}{2} n_i(n_i-1) + U n_i m_i \right] \label{hamiltonian} \\
&& + \sum_i \left[-\mu^\textrm{b}_i n_i -\mu^\textrm{f}_i m_i \right] \nonumber
\end{eqnarray}
where $b_{i}$ and $f_{i}$ are bosonic and fermionic annihilation operators of a 
particle in the $i$-th site and $n_i= b^{\dagger}_i b_i$, $m_i= f^{\dagger}_i f_i$ 
are the corresponding on-site number operators.
The FBH model describes:
(i) nearest neighbor boson (fermion) hopping, with an associated negative energy,
$-J_\textrm{b}$ ($-J_\textrm{f}$);
(ii) on-site boson-boson interactions with an energy $V$, which is supposed to be positive ({\it i.e.} repulsive) in the reminder of the paper;
(iii) on-site boson-fermion interactions with an energy $U$, which is positive (negative) for repulsive (attractive) interactions;
(iv) on-site energy due to interactions with a possibly inhomogeneous potential, with energies $-\mu^\textrm{b}_i$ and $-\mu^\textrm{f}_i$. 
Eventually, $-\mu^\textrm{b}_i$ and $-\mu^\textrm{f}_i$ also contain 
the chemical potentials in grand canonical description.

In the following, we investigate the properties of the strong coupling regime, 
{\it i.e.} $V,U \gg J_\textrm{b,f}$ and
we derive a one-species effective Hamiltonian using a perturbative
development up to second order in $J_\textrm{b,f}/V$.

\subsection{Zeroth-order perturbation: formation of fermion composites}
\label{sec:composites}
In the limit of a vanishing hopping ($J_\textrm{b}=J_\textrm{f}=0$) with 
finite repulsive boson-boson interaction $V$,
and in the absence of interactions between bosons and fermions ($U=0$), 
the bosons are pinned in the lattice sites with exactly 
$\tilde{n}_i=\lceil\tilde{\mu}_i^\textrm{b}\rceil+1$ bosons per site,
where $\tilde{\mu}^\textrm{b}=\mu^\textrm{b}/V$ and 
$\lceil x \rceil$ denotes the integer part of $x$. 
For simplicity, we assume $\tilde{n}_i=1$ for all sites so that 
the boson system is in the Mott insulator (MI) phase with $1$ boson per site.
In contrast, the fermions can be in any set of Wannier states, since for 
a vanishing tunneling, 
the energy of the system does not depend on the distribution of the fermions in different lattice sites for the homogeneous optical lattice and will be very similar for the different repartitions of fermions in the case of small disorder.

Assume now that the boson-fermion interaction is turned on positive ($U>0$)
and define $\alpha=U/V$. The presence of
a fermion in site $i$ may expel $s = 0 \textrm{~or~} 1$ boson
depending on the interaction strength.
The on-site energy gain in expelling $s$ boson from site $i$ is
$\Delta E_i = \frac{V}{2}s(s-1) - U s + \mu_i^\textrm{b} s$.
Minimizing $\Delta E_i$ versus $s$, we find that
$s=\left\lceil\alpha-\tilde{\mu}_i^\textrm{b}\right\rceil+1$ that we assume to 
be $1$ in all sites.
Within the occupation number basis, excitations correspond to having $1$ boson 
in a site with a fermion, instead of $0$ boson and, therefore, the corresponding 
excitation energy is $\sim U$.
In the following, we assume that the temperature is smaller than $U$ so that 
the populations of the
above mentioned excitations can be neglected. 

Summarizing the discussion above, we have assumed that 
$0 < \mu_i^\textrm{b}<V$ (so that $\tilde{n}_i=1$), 
$U-V<\mu_i^\textrm{b}<U$ (so that in the lowest energy states, each lattice site 
is populated by either one boson or one fermion but never $0$ or $2$ particles),
and
$k_\textrm{B}T<U$ (so that only these lowest energy states are significantly populated).
The physics of our Fermi-Bose
mixture can thus be regarded as the one of composite particles 
made of one fermion plus one bosonic hole. 
Notice that the bosonic hole is created because the initially present boson is expelled from the site by the fermion, due to the repulsive interactions between bosons and fermions.
Within the picture of composites, a lattice site is either populated by one composite (i.e. by one fermion plus bosonic hole) or free of composite (i.e. populated by one boson).
In particular, the vacuum state 
corresponds to the MI phase with $1$ boson per site.
The annihilation and creation operators of the composites are 
\cite{lewenstein2004}:
\begin{eqnarray}
F_i & = & b_i^\dagger f_i \mathcal{P} \label{FI} \\
F_i^\dagger & = &  \mathcal{P} f_i^\dagger b_i  \label{FIdag} 
\end{eqnarray}
where $\mathcal{P}$ is the projector onto the sub-Hilbert space of the composites.
It is straightforward to show that $F_i$ and $F_i^\dagger$ are fermionic operators.

\subsection{Second-order perturbation: effective Hamiltonian}
\label{sec:effective}
The sub-Hilbert space of the composites is 
$[N! / (N_\textrm{f})! (N - N_\textrm{f})!]$-dimensional,
a number that corresponds to the number of possibilities to
distribute the $N_\textrm{f}$ fermions into the $N$ lattice sites.
We assume now that the tunneling rates $J_\textrm{b}$ and $J_\textrm{f}$ are
small but finite. 
For weak enough disorder, one can assume site independent tunneling rates for bosons and fermions
\cite{damski2003}.
Using second order projection perturbation theory \cite{cohen}, 
we derive an effective Hamiltonian
for the fermionic composites by means of a unitary transform applied to the total
Hamiltonian (\ref{hamiltonian}).
The general expression of the effective Hamiltonian can be written as:
\begin{eqnarray}
&&\langle out | H_\textrm{eff}|in\rangle =
\langle out |H_{0}|in\rangle + \langle out | \mathcal{P}H_{int}\mathcal{P}|in\rangle \label{hamil} \\
&&-\frac{1}{2}\langle out|
\mathcal{P}H_{int}\mathcal{Q}\left[\frac{1}{H_0-E_{in}}+\frac{1}{H_0-E_{out}}\right]
\mathcal{Q}H_{int}\mathcal{P}|in\rangle.\nonumber
\end{eqnarray}
where $\mathcal{Q}=1-\mathcal{P}$ is the complement of the projection operator
$\mathcal{P}$.
As second order theory can only connect states that differ on one set of two adjacent 
sites, $H_\textrm{eff}$ can only contain nearest-neighbor hopping and interactions as well as
on-site energies $\overline{\mu}_i$ \cite{sanpera2004}:
\begin{equation}
H_\textrm{eff}=\sum_{\langle i,j \rangle} \left[ -d_{i,j} F^{\dagger}_i F_j + h.c.\right] + \sum_{\langle i,j \rangle} K_{i,j} M_i M_j + \sum_i \overline{\mu}_i M_i
\label{Heffinhom}
\end{equation}
where $F_i$ and $F^{\dagger}_i$ are defined in Eqs.~(\ref{FI}-\ref{FIdag}) and $M_i=F^{\dagger}_i F_i$ is the composite number operator in lattice site $i$. 
Although the general form of Hamiltonian~(\ref{Heffinhom}) is universal for all types of 
composites, the explicit calculation of the
coefficients $d_{i,j}$, $K_{i,j}$ and $\overline{\mu}_i$ depends on the concrete 
type of composites that can be created for different values and sign of alpha \cite{ahufinger2005}.
The nearest neighbor hopping for
the composites is described by $-d_{i,j}$ and the nearest neighbor 
composite-composite interaction
is given by $K_{i,j}$, which may be repulsive ($>0$) or attractive ($<0$). 
This effective model is
equivalent to that of spinless interacting fermions \cite{auerbach1994}. 

In the situation under consideration, each site contains
either one boson or one fermion. 
Therefore, a fermion jump from site $i$ to site $j$
can only occur if the boson that was initially in site $j$ jumps back to site $i$
into the hole the fermion leaves behind.
 Therefore, the number operator for fermions and bosons are related to the number
operator of composites, {\it i.e.} $M_i=m_i=1-n_i$. 
From Eq.~(\ref{hamil}), we find
\begin{eqnarray}
d_{ij} &=& \frac{J_\textrm{b}J_\textrm{f}}{V}\left(\frac{\alpha}{\alpha^2-(\Delta^\textrm{b}_{ij})^2}
+\frac{\alpha}{\alpha^2-(\Delta^\textrm{f}_{ij})^2}\right)\label{dij2} \\
K_{ij} & = & -\frac{J_\textrm{b}^2}{V}\Bigg(\frac{4}{1-(\Delta^\textrm{b}_{ij})^2}
-\frac{2\alpha}{\alpha^2-(\Delta^\textrm{b}_{ij})^2}\Bigg) \nonumber \\
&& -\frac{J_\textrm{f}^2}{V}\Bigg(\frac{2\alpha}{\alpha^2-(\Delta^\textrm{f}_{ij})^2}\Bigg) \label{kij2} \\
\overline{\mu}_i &=& \mu_i^{\textrm{b}}-\mu_i^{\textrm{f}}
+ \frac{J_\textrm{b}^2}{V} \sum_{\langle i,j \rangle}
\left[ 
{\frac{4}{1-(\Delta_{ij}^{\textrm{b}})^2}}-
{\frac{1}{\alpha-\Delta_{ij}^{\textrm{b}}}}
\right] \nonumber \\
&& - \frac{J_\textrm{f}^2}{V} \sum_{\langle i,j \rangle}
\left[
{\frac{1}{\alpha+\Delta_{ij}^{\textrm{f}}}}
\right] \label{muij2}
\end{eqnarray}
with $\Delta_{ij}^\textrm{f,b}=\tilde{\mu}_i^\textrm{f,b}-\tilde{\mu}_j^\textrm{f,b}$
measures the inhomogeneity of the lattice sites.
Here, $\langle i,j \rangle$ represents all sites $j$ adjacent to site $i$. 
The coupling parameters in Hamiltonian~(\ref{Heffinhom}) depend on all parameters
($J_\textrm{b}$, $J_\textrm{f}$, $U$, $V$, $\mu_i^\textrm{b}$, $\mu_i^\textrm{f}$)
of the FBH model~(\ref{hamiltonian}) in a complex fashion.
For example, the hopping amplitudes $d_{i,j}$ depend on
disorder but always remain positive.
In contrast, the interaction term $K_{i,j}$ can be positive or negative.
All terms $d_{i,j}$, $K_{i,j}$, $\overline{\mu}_i$ are random due to the
inhomogeneous on-site energies $\mu_i^\textrm{b}$ and $\mu_i^\textrm{f}$ of 
the bosons and the fermions respectively.
It is thus clear that the {\it qualitative character of interactions}
may be {\it controlled by the inhomogeneities} \cite{sanpera2004}. 

\begin{figure}
\includegraphics[width=1.0\linewidth]{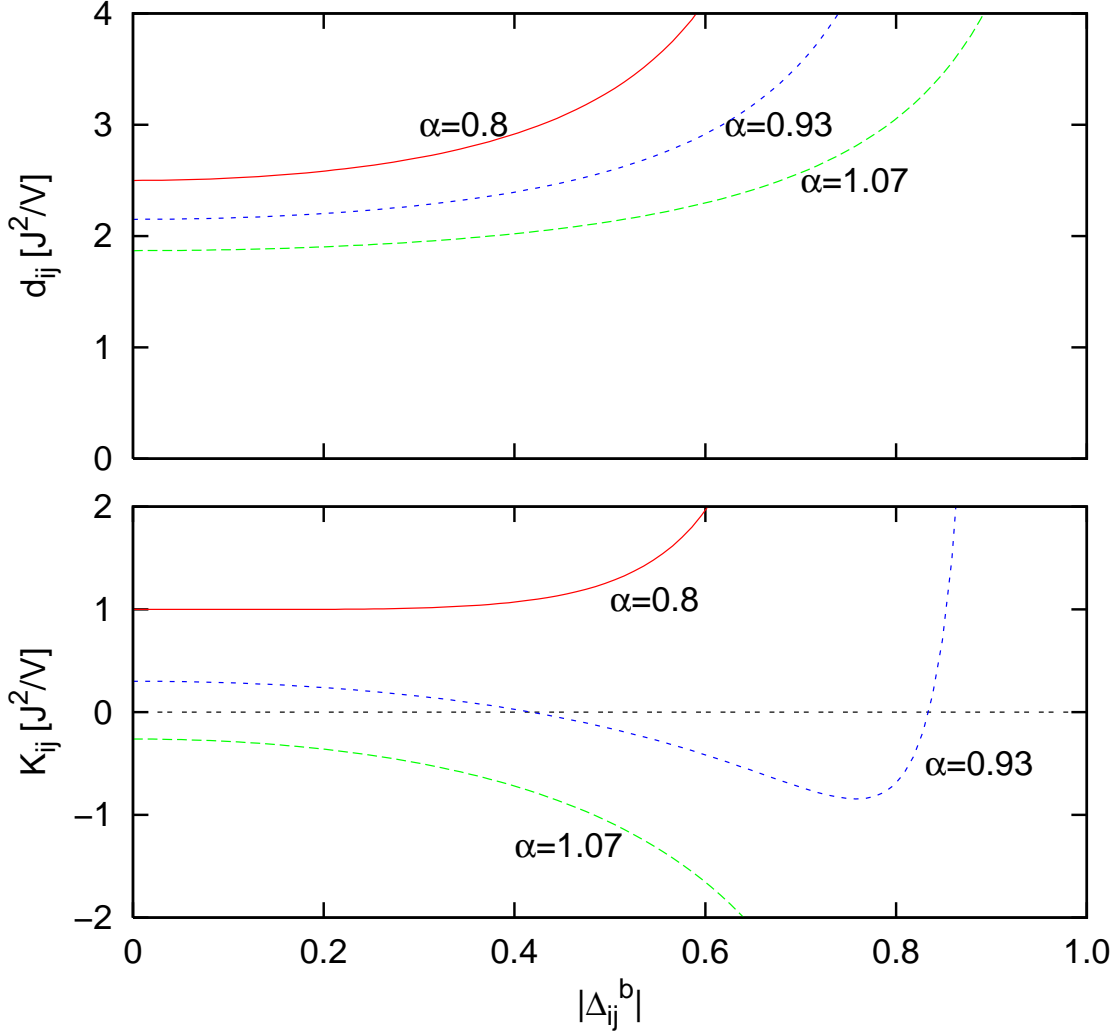}
\caption{(color online) 
Tunneling, $d_{ij}$, and nearest neighbor interactions $K_{ij}$ between the
fermion composites as a function of the disorder of bosons $\Delta_{ij}^\textrm{b}$
for various boson-fermion interactions $\alpha$. 
The disorder for fermions is assumed to vanish here ($\mu_i^\textrm{f}=0$).
\label{fig:sitesB}}
\end{figure}

For example, consider the case where disorder only applies to the bosons 
($\mu_i^\textrm{f}=cst$, uniform) \cite{note:experiments}.
The corresponding coupling parameters are plotted in Fig.~\ref{fig:sitesB}. As mentioned
above, the hopping amplitudes $d_{ij}$ are always positive, although may vary quite significantly with disorder, especially when $\Delta_{ij}^\textrm{b}\simeq\alpha$.
As shown in Fig.~\ref{fig:sitesB}, for $\alpha>1$, $K_{ij}\le 0$ and we deal with attractive (although random) interactions. For $\alpha<1$, $K_{ij}\ge 0$ and the interactions between composites are repulsive. For $\alpha<1$, but close to 1, $K_{ij}$ might take positive or negative values for $\Delta_{ij}^\textrm{b}$ small or
$\Delta_{ij}^\textrm{b}\simeq \alpha$. 

In the following two sections, we investigate the ground state properties of the
Fermi-Bose mixture (or equivalently that of the fermion composites) in the 
presence of diagonal disorder. We distinguish two limiting cases. The first
corresponds to small disorder while the second corresponds to $\Delta_{i,j}^{(b)} \simeq \alpha$
that interestingly maps to a spin glass problem.

\section{Quantum phase diagrams of Fermi-Bose mixtures in weakly disordered optical lattices}
\label{sec:weak}
The first limiting case corresponding to weak disorder ($\Delta_{i,j}^\textrm{(f,b)} \ll 1, \alpha$)
is investigated in this section.
For the sake of simplicity, we use 
$J_\textrm{b}=J_\textrm{f}=J$.
In this limit, the contributions of the inhomogeneities
$\Delta_{ij}^\textrm{b,f}$ to the tunneling $d_{ij}$ and to 
the interaction $K_{ij}$ terms in the effective composite 
Hamiltonian~(\ref{Heffinhom}) can be neglected ($d_{ij}\simeq d$ and $K_{ij}\simeq K$)
and we keep only the leading disorder contribution in 
$\overline\mu_i$ [first term in Eq.~(\ref{muij2})]. 
Note, that the latter contribution is particularly relevant in 1D and 2D
leading to Anderson localization of single particles \cite{gang1979}.
We will first describe the non-disordered case and then we will discuss the effect of weak
disorder. 

The analysis reported below is supported by mean-field numerical calculations
(for details about the numerical method, see in Ref.~\cite{ahufinger2005}).
We consider a 2D optical lattice with $N=100$ sites with 
$N_\textrm{b}=60$ bosons,
$N_\textrm{f}=40$ fermions and
$J_\textrm{b}/V=J_\textrm{f}/V=0.02$
to compute the ground state of the system in the presence of a very shallow harmonic
trapping potential ($\mu_i^\textrm{b,f} = \omega^\textrm{b,f}\times l(i)^2$, 
where $l(i)$ is the distance from site $i$ to the center in cell size units)
with eventually different amplitudes for bosons and fermions. 
This simulates optical or magnetic trapping which turns out to be hardly avoidable in 
current experiments on ultracold atoms.
In addition, it breaks the equivalence of all lattice sites and makes more 
obvious the different phases that one can achieve (see below).
In the numerics, we have used 
$\omega^\textrm{b}=10^{-7}$ and $\omega^\textrm{f}=5\times 10^{-7}$.

In the absence of interactions between the bosons and the fermions ($\alpha=0$), 
the bosons are well inside the MI phase with $\tilde n=1$ boson per site in the center of the trap \cite{fisher1989,jaksch1998}. 
Besides, the non-interacting fermions are delocalized and due to the very small
value of $\omega^\textrm{f}$ they do not feel significantly the confining trap as 
shown in Fig.~\ref{fig:no_disorder}(a).

\subsection{Quantum phases in non-disordered optical lattices}
In the absence of disorder, the physics of the Fermi-Bose mixture is mainly determined by the 
ratio $K/d$ and the sign of $K$ where $K$ and $d$ are site independent parameters. 
Once the fermion composites are created 
($\alpha>\tilde \mu^\textrm{b}$, see section~\ref{sec:composites}),
we have $K/d=-2(\alpha-1)$ as a result of
Eqs.~(\ref{dij2}-\ref{kij2}) with 
$J_\textrm{b}=J_\textrm{f}=J$ and $\Delta_{ij}=0$. 
We now
discuss the quantum phases that are accessible depending on the control parameter $\alpha$
\cite{note:alpha}.

\begin{figure}[ht!]
\includegraphics[width=1.0\linewidth]{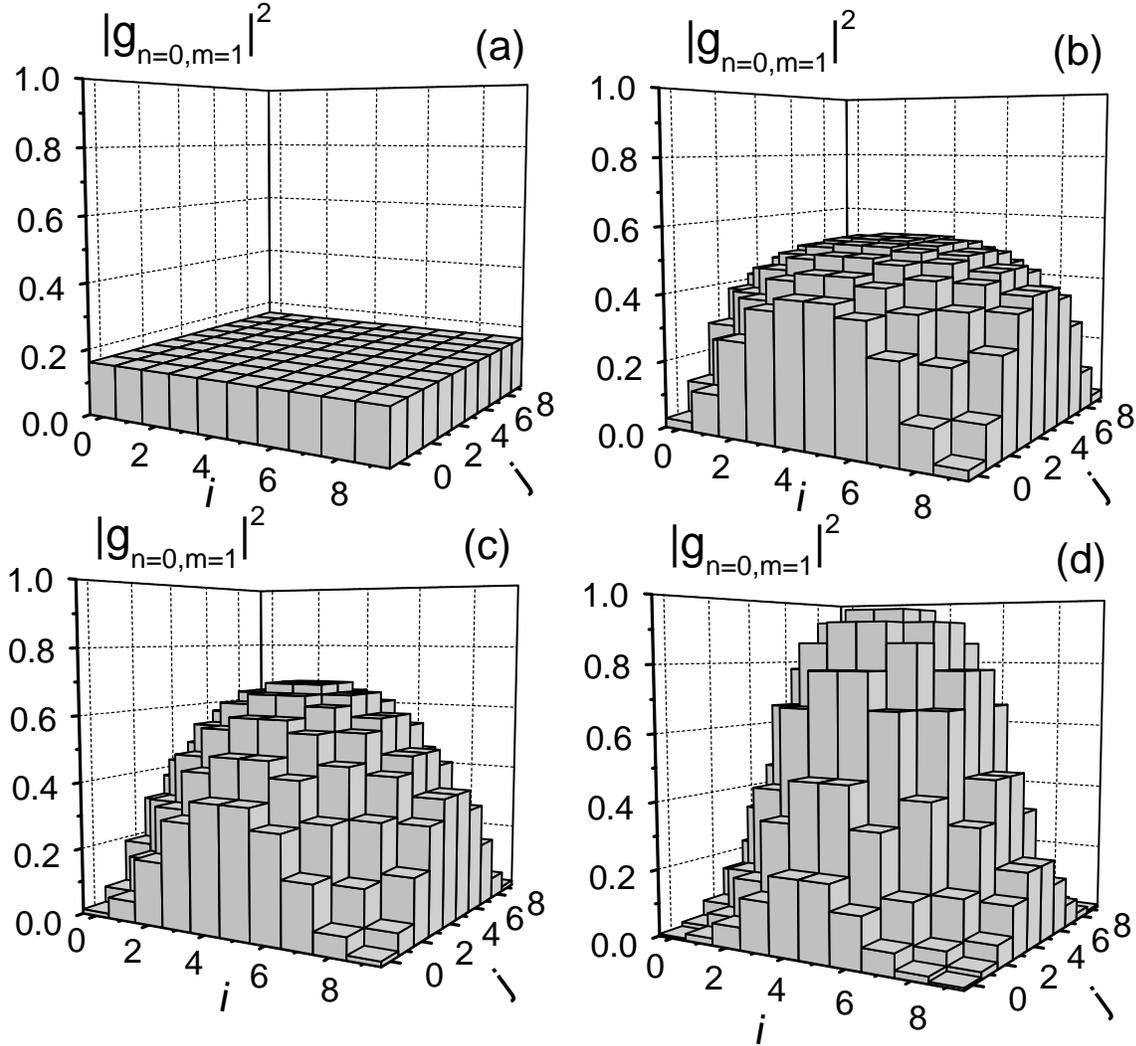}
\caption{Local on-site probability of finding one fermion and zero boson at each lattice site for
the $N=100$ lattice sites in a Fermi-Bose mixture with $N_\textrm{b}=60$, $N_\textrm{f}=40$, $J_\textrm{b}/V=J_\textrm{f}/V=0.02$
and in the presence of harmonic traps for bosons and fermions characterized by $\omega_\textrm{b}=10^{-7}$ and $\omega_\textrm{f}=5\times 10^{-7}$, respectively.
The interaction between fermions and bosons is 
(a) $\alpha=0$ [independent bosonic MI and Fermi gas],
(b) $\alpha=0.5$ [Fermi liquid], 
(c) $\alpha=1$ [ideal Fermi gas] and 
(d) $\alpha=10$ [fermionic insulator domain].}
\label{fig:no_disorder}
\end{figure}

\paragraph{}
For $\tilde{\mu}^\textrm{b}<\alpha<1$, 
the interactions between composite fermions are repulsive and of
the same order of magnitude as the tunneling ($K \sim d$).
Therefore, the system enters an interacting {\it Fermi liquid} quantum phase
[see Fig.~\ref{fig:no_disorder}(b)]. 
The fermions are delocalized over the entire lattice but populate preferably
the center of the confining trap. Small repulsive composite interactions
tend to flatten the density profile compared to that of non-interacting composites 
(see Fig.~\ref{fig:no_disorder} and text below).

\paragraph{}
For $\alpha \simeq 1$, although the interactions between the bosons and the fermions
are large, the interactions between the fermion composites vanish and 
the system shows up properties of an {\it ideal Fermi gas}
[see Fig.~\ref{fig:no_disorder}(c)]. 
Again, the fermions are delocalized and their distribution follows the harmonic confinement.

\paragraph{}
Growing further the repulsive interactions between bosons and fermions,
the interactions between the fermion composites become attractive.
For  $1<\alpha<2$, one expects the system to be a weakly interacting {\it superfluid}, whereas
for $\alpha>2$ a {\it fermionic insulator domain} phase is predicted
[see Fig.~\ref{fig:no_disorder}(d)]. In this case, the fermions are pinned in
the lattice sites. They tend to merge because of site-to-site attractive interactions
and populate the center of the trap.

Notice that contrary to the bare fermions, the composite fermions
are significantly affected by the harmonic trapping potential. This is because the
coupling parameters of the composite effective Hamiltonian~(\ref{Heffinhom})
are much smaller than the coupling parameters of the bare fermion-boson
Hubbard Hamiltonian~(\ref{hamiltonian}). Therefore the harmonic potential
is able to compete with tunneling and interactions for the composites \cite{ahufinger2005}.

\subsection{Quantum phases in disordered optical lattices}
\begin{figure}[h!]
\includegraphics[width=1.0\linewidth]{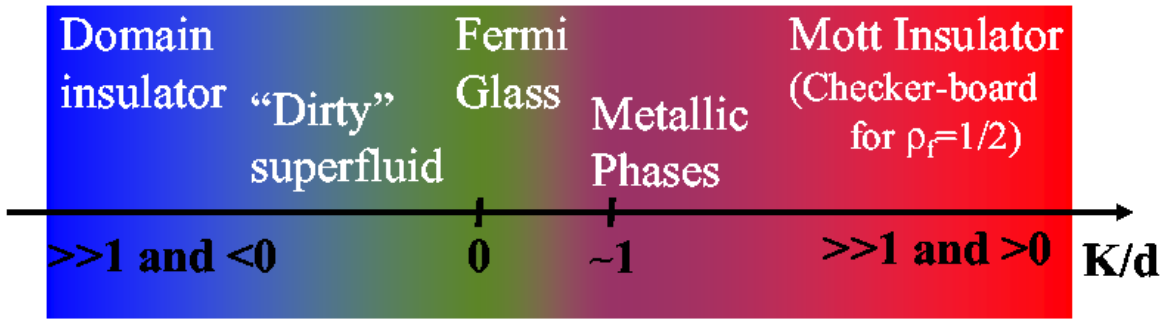}
\caption{(color online) Schematic phase diagram of fermionic composites
for small disorder ($\Delta_{ij}^\textrm{b} \ll 1, \alpha$) 
as a function of the ratio between nearest neighbor interactions and 
tunneling of the composites ($K/d$).
\label{fig:sitesBbis}
}
\end{figure}

We now assume that small on-site inhomogeneities are present and we investigate
the effect of disorder on the quantum phase diagram of the system depending on the 
parameters of the effective Hamiltonian~(\ref{Heffinhom}). 
Fig.~\ref{fig:sitesBbis} shows a schematic representation of expected disordered 
phases of the fermionic composites for small disorder.

\paragraph{} For $|K|\ll d$, the system is in the {\it Fermi glass} phase,
{\it i.e.} Anderson localized (and many-body corrected) single particle
states are occupied according to the Fermi-Dirac statistics \cite{fermiglass}.
The effect of disorder is to localize the fermions preferably into the
deepest sites. 

\paragraph{} For repulsive and large composite interactions ($K>0$ and $K \gg d$),
the ground state is a Fermi MI phase and the composite fermions are pinned at large 
filling factors preferably into the deepest wells. 
At half filling factor ($\rho_\textrm{f}=1/2$), one expects the ground state to form a {\it checker-board}, 
i.e. the lattice sites are alternatively empty and populated by one composite.

\paragraph{} For large attractive composite interactions ($K<0$ and $|K| \gg d$), 
the fermions form a {\it domain insulator} which average position results from
the competition between the random and the confining potential.

\paragraph{} 
Finally, for intermediate values of $K/d$, with $K>0$, {\it delocalized metallic} 
phases  with enhanced persistent currents are possible \cite{metalglass}.
Similarly, for attractive interactions ($K<0$) and  $|K|< d$ one expects 
a competition between pairing of fermions and disorder 
{\it i.e.} a {\it dirty superfluid} phase.

Further information is provided by numerical computations that we present now. 
We consider on-site random inhomogeneities for the bosons
$\mu_i^\textrm{b}$. We start from a non-disordered phase
[$\Delta (t=0)=0$] and we slowly increase the
standard deviation of the disorder 
$\sqrt{\langle (\tilde{\mu}_i^{\textrm{b}})^2 \rangle - (\langle\tilde{\mu}_i^{\textrm{b}}\rangle)^2 }=\Delta(t)$ from $0$ to its final value $\Delta$.

\begin{figure}
\includegraphics[width=0.9\linewidth]{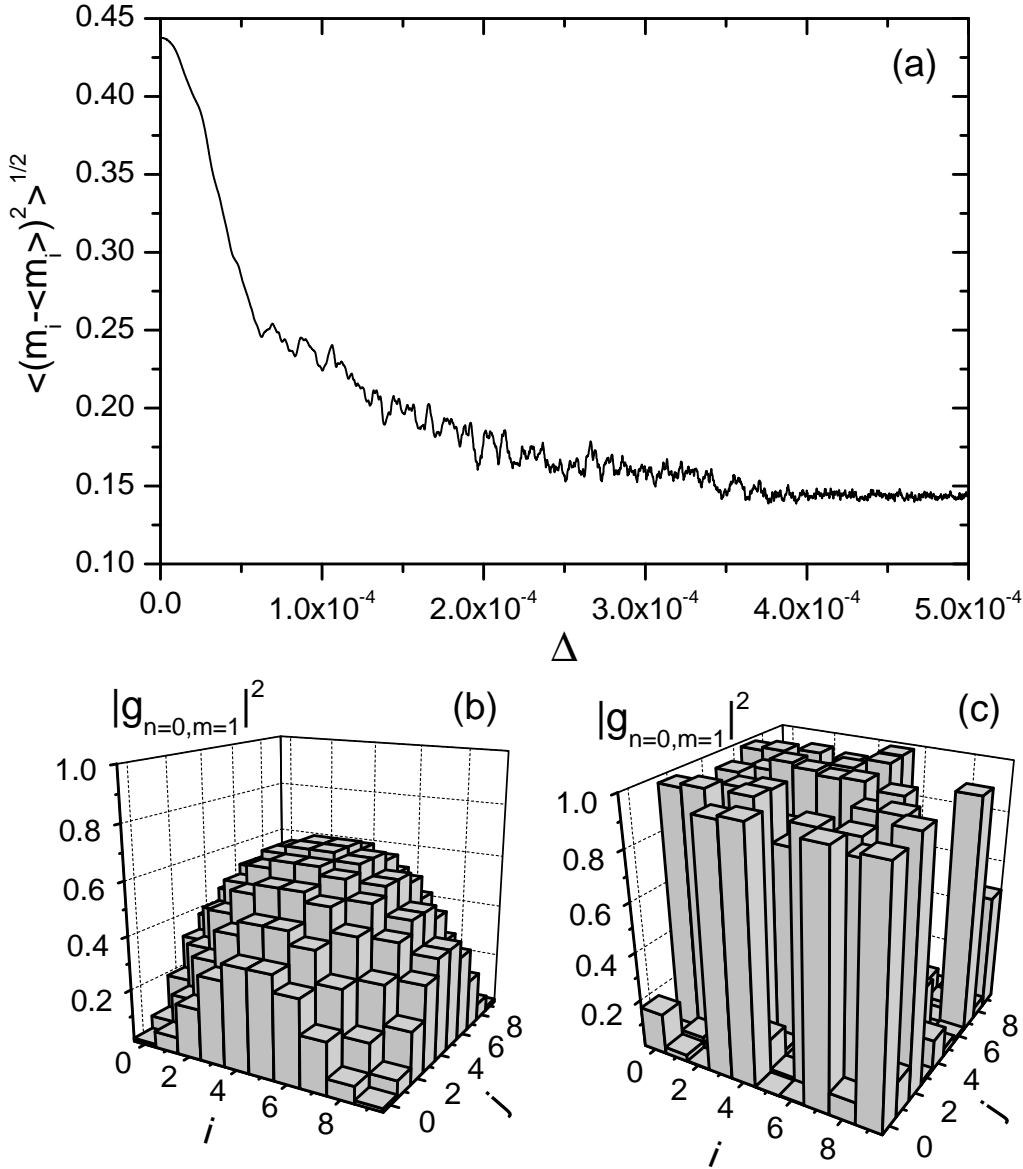}
\caption{Dynamical crossover from the Fermi gas to the Fermi glass phases. The parameters are the same as in
Fig.~\ref{fig:no_disorder}(c). (a) Variance of the number of fermions per lattice site as a function of the amplitude of the disorder $\Delta$. (b) Probability of having one composite (one fermion and zero boson) at each lattice site for the M sites
in the absence of disorder and (c) after ramping up adiabatically diagonal disorder with amplitude $\Delta=5\times 10^{-4}$.}\label{fig:dis_Fermi_liquid}
\end{figure}

We first study the transition from a (composite) {\it Fermi gas} in the absence of
disorder [see Fig.~\ref{fig:dis_Fermi_liquid}(b)] to a (composite) {\it Fermi glass} [see Fig.~\ref{fig:dis_Fermi_liquid}(c)]. Initially [$\Delta(t)=0$], the composite fermions are delocalized although 
confined near the center of the effective harmonic potential 
[$(\omega_\textrm{f}-\omega_{b})\times l(i)^2$]. The local populations
fluctuate around  $\langle m_i \rangle \simeq 0.4$
with a standard deviation
$\sqrt{\langle (m_i - \langle m_i \rangle)^2 \rangle} \simeq 0.43$. 
Increasing the amplitude of disorder, 
the fluctuations of $m_i$ decreases as shown in Fig.~\ref{fig:dis_Fermi_liquid}(a). 
This indicates that the composites localize more and more in the 
lattice sites to form a Fermi glass. 
For $\Delta=5 \times 10^{-4}$, the composite fermions are pinned in random sites 
as shown in Fig.~\ref{fig:dis_Fermi_liquid}(c).
As expected, the $N_\textrm{f}$ composite fermions populate the $N_\textrm{f}$ 
sites with minimal $\tilde{\mu}_i^\textrm{b}$.

\begin{figure}
\includegraphics[width=0.9\linewidth]{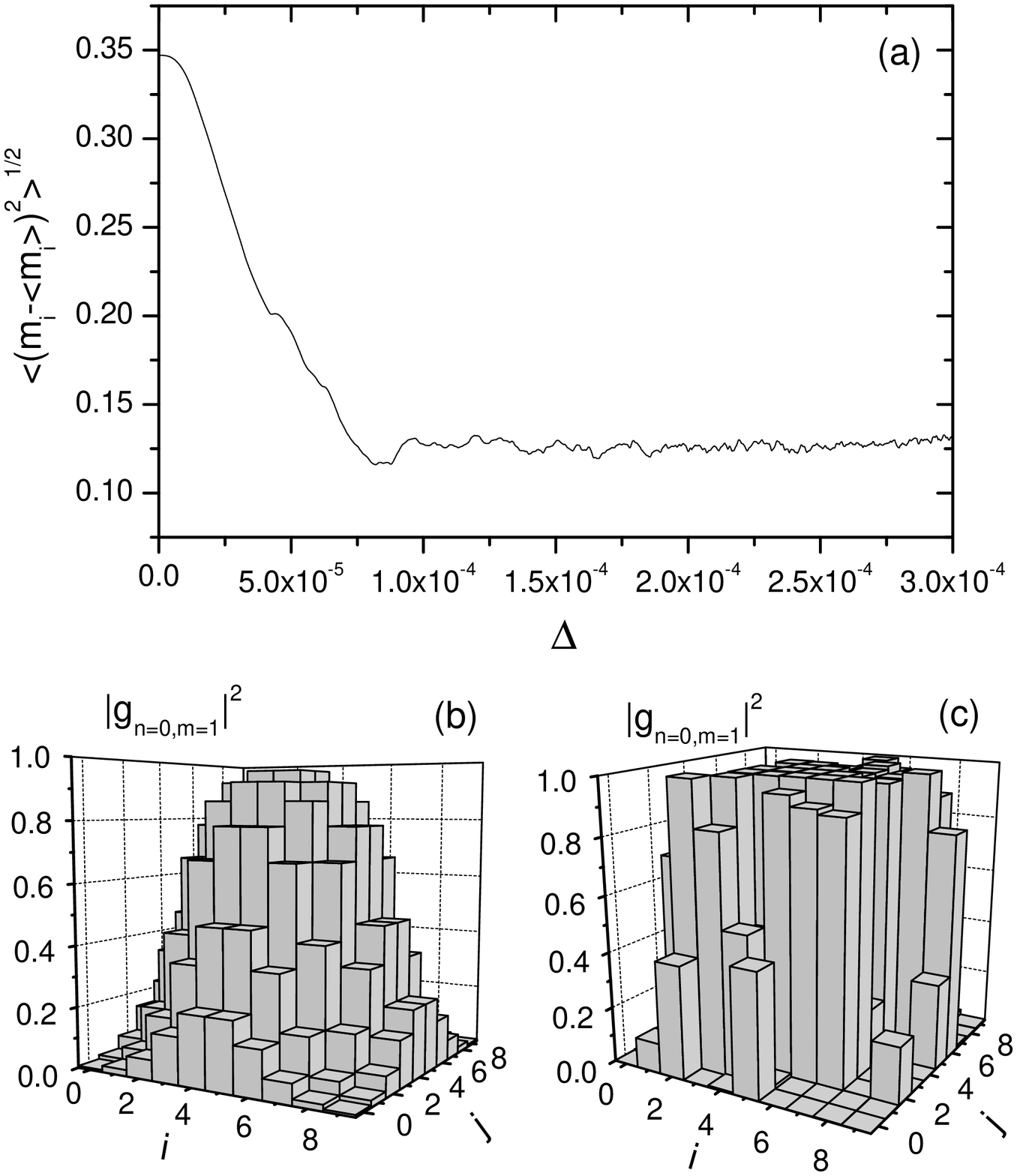}
\caption{Dynamical crossover from the fermionic domain insulator to a
disordered insulating phase. The parameters correspond to
Fig.~\ref{fig:no_disorder}(d). (a) Variance of the number of fermions per lattice site as
a function of the amplitude of the disorder. (b) Probability of having one composite (one fermion and zero boson) in
 each lattice site in the absence of disorder and (c) after ramping up adiabatically
the disorder with amplitude $\Delta=3\times 10^{-4}$.}\label{fig:dis_domains}
\end{figure}

We now turn to the transition from a {\it Fermi insulator domain} phase [see Fig.~\ref{fig:dis_domains}(b)] to a {\it disordered insulating phase} while slowly increasing 
the amplitude of the disorder. For the Fermi insulator domain, the composite 
fermions are pinned near the center of the harmonic trap and surrounded by a 
ring of delocalized fermions which results in finite fluctuations on the fermion occupation
number ($\sqrt{\langle (m_i - \langle m_i \rangle)^2 \rangle} \simeq 0.35$). 
As shown in Fig.~\ref{fig:dis_domains}(a), while ramping up the amplitude of disorder, 
the fluctuations decrease down to $\sqrt{\langle (m_i - \langle m_i \rangle)^2 \rangle} \simeq 0.13$ for $\Delta>10^{-4}$ showing that the composite fermions are pinned in different lattice sites.
This can be seen in the plot the site population of the composite fermions presented in Fig.~\ref{fig:dis_domains}(c). 
Contrary to what happens for the transition from the Fermi gas to Fermi glass, the 
composites mostly populate the central part of the confining
potential. This is because (i) the attractive interaction between
composites is of the order of $K \simeq -1.4\times 10^{-3}$ and competes with disorder
($\Delta = 3\times 10^{-4}$) and (ii) because tunneling is small 
($d\simeq 8\times 10^{-5}$) so the fermions can hardly move during the ramp of
disorder.

\section{The spin glass limit}
\label{sec:spinglass}

\subsection{From composite Hamiltonians to spin glasses models}
The second limit corresponds to a case where the interactions between fermions and bosons are
of the order of, but slightly smaller than the interactions between the bosons ($\alpha \sim 1$
and $\alpha <1$). As shown in Fig.~\ref{fig:sitesB}, the effective interaction term $K_{i,j}$ has 
a zero at large inhomogeneities $|\Delta_{i,j}^\textrm{b}|$ and varies strongly with 
$|\Delta_{i,j}^\textrm{b}|$ reaching both positive to negative values.
Such a situation is accessible for ultracold atomic systems
using a superlattice with a spatial period twice as large
as the lattice spacing plus a random potential, both acting on the bosons. 
The interaction with this superlattice results in an alternatively
positive and negative additional on-site energy of the bosons, and
whose amplitude is controlled by the intensity of the superlattice. In particular, one can set
$|\Delta_{i,j}| \simeq \alpha$. An additional weak random potential introduces disorder. 

Due to the random on-site effective energy $\overline{\mu}_i$, 
the effective tunneling becomes non-resonant and can be neglected 
in first approximation while $K_{i,j}$ is random
with a given average (eventually zero) and strong fluctuations 
from positive to negative values.  
The effective Hamiltonian~(\ref{Heffinhom}) then reduces to 
\begin{equation}
H_\textrm{\tiny EA}=\frac{1}{4}\sum_{\left\langle ij\right\rangle }
K_{ij}s_is_j + \frac{1}{2}\sum_{i}\overline\mu_i s_i ,
\label{HamiltonianSG}
\end{equation}
where we have introduced $s_i=2M_i-1=\pm 1$. Interpreting $s_i$ as classical Ising spins \cite{note:spins},
this Hamiltonian is equivalent to the well known Edwards-Anderson model \cite{edwards1975}. 
This describes {\it spin glasses}, i.e., a Ising model with random positive (anti-ferromagnetic) 
or negative (ferromagnetic) exchange terms $K_{i,j}$. Our system however differs from the usual
Edwards-Anderson spin glass model as (i) it has a random magnetic term $\overline\mu_i$ and 
(ii) the average magnetization per site $m=2N_\textrm{f}/N-1$ is fixed by the total number
of fermions in the lattice.
It however shares basic characteristics with spin glasses as being a
spin Hamiltonian with random spin exchange terms $K_{ij}$. 
In particular, this provides bond frustration,
which
is essential for the appearance of the spin glass phase and turns out to introduce
severe difficulties for analytical and numerical analyses. 
We thus think that experiments with ultracold atoms can provide a useful {\it quantum
simulator} to address challenging questions related to spin glasses such as the nature of the 
ordering of its ground- and possibly metastable states \cite{mezard1987,huse,stein2003},
broken symmetry and dynamics of spin glasses \cite{sachdev1999,georges2001}.

In the following two sections, we outline some general properties of spin glasses and then we
apply the {\it replica method} under the constraint of a fixed magnetization and argue that
this preserves the occurrence of a symmetry breaking characteristics of spin glasses in
the M\'ezard-Parisi theory \cite{mezard1987}.

\subsection{Generalities on spin glasses}
Consider a spin glass at finite temperature with a random exchange term $K_{ij}$ with
average $\overline{K}$ and variance $\Delta K$. The magnetization is characterized
by two order parameters: (i) $m=\overline{\langle s_i \rangle}$, the average magnetization
per site and (ii) $q_\textrm{\tiny EA}=\overline{\langle s_i \rangle^2}$, the Edwards-Anderson
parameter, where $\overline{~\cdot~}$ denotes the average over disorder while 
$\langle\cdot\rangle$ represents the thermodynamics average. 
It is clear that $m \neq 0$ signals a long-range magnetic order while 
$q_\textrm{\tiny EA} \neq 0$ signals a local magnetization that may vary from site to site
and from one configuration of quenched disorder to another.

\begin{figure}
\includegraphics[width=0.8\linewidth]{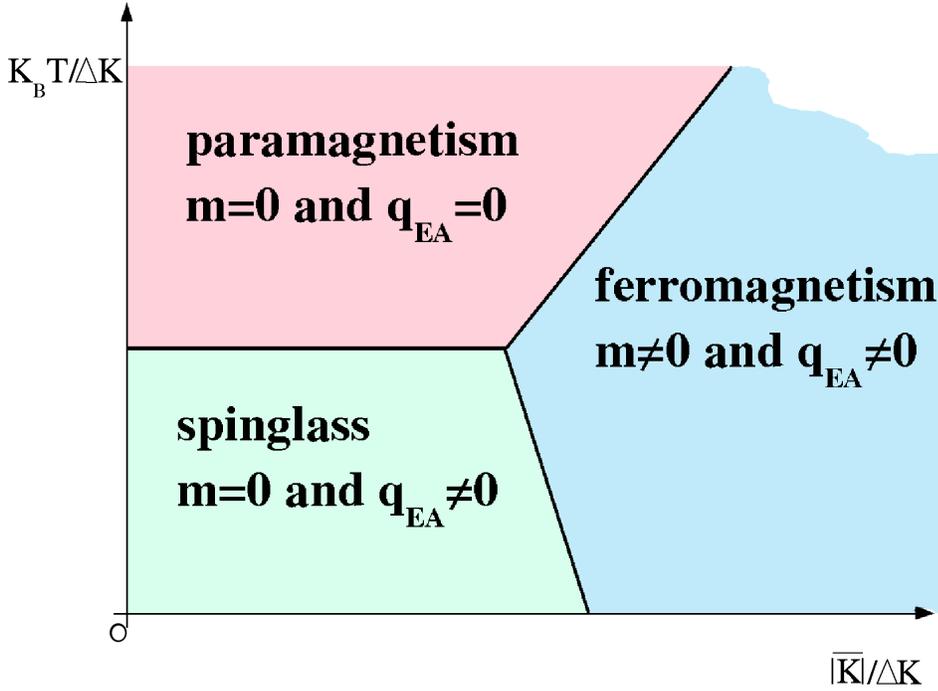}
\caption{Schematic phase diagram of a spinglass.}
\label{fig:spinglass}
\end{figure}

Earlier experimental studies have identified three magnetic phases as schematically
represented in Fig.~\ref{fig:spinglass} \cite{sherrington1998}:
(i) At high temperature and small average spin exchange $\overline{K}$, one finds
a {\it paramagnetic phase} characterized by $m=q_\textrm{\tiny EA}=0$.
(ii) For $\overline{K}>0$ and large, one has a {\it ferromagnet} with
$m \neq 0$ and $q_\textrm{\tiny EA} \neq 0$.
(iii) For weak $\overline{K}$ and small temperatures, a {\it spin glass} phase appears
with $m = 0$ but $q_\textrm{\tiny EA} \neq 0$. This signals that the local magnetization is
frozen but that disorder prevents a long-range magnetic order.

As pointed out before, the physics of spin glasses still opens challenging questions. In
particular, there are competing theories that predict different natures of the magnetic
order. 

{\it The 'droplet' picture} is a phenomenological theory based on numerical and scaling arguments. It predicts that there are two ground states related by spin-flip symmetry and that low-lying excitations are domains with fractal boundaries (the droplets) with all spins 
inversed compared to the groundstate. This theory is supported by numerical
computations and is believed to be valid in short-range spin glass
models such as the Edwards-Anderson 
\cite{edwards1975}.

{\it The M\'ezard-Parisi picture} predicts a large number of low-energy states with very
similar energies. This leads in particular to 
{\it disorder-induced quantum frustration}.
This theory is a mean-field, based on the {\it replica method} which 
is a special trick introduced to compute the non-trivial average over disorder of the free 
energy functional. The M\'ezard-Parisi theory is formulated in long-range spin glasses such as
the Sherrington-Kirkpartick model
\begin{equation}
H_\textrm{S-K}=\frac{1}{4}\sum_{\left( ij\right) }
K_{ij}s_is_j + \frac{1}{2}\sum_{i}\overline\mu_i s_i ,
\label{HamiltonianSG2}
\end{equation}
that differs from the Edwards-Anderson model~(\ref{HamiltonianSG}) by the long-range spin 
exchange 
[$(ij)$ denotes here the sum over {\it all} pairs of lattice sites, either neighbors 
or not]. The applicability of the M\'ezard-Parisi picture to short-range spin glasses
is still controversial.

\subsection{The replica-symmetric solution for fixed magnetization}
As the disorder is quenched, one must average over disorder the free energy density,
$\overline{f}=-\overline{\ln Z}/k_\textrm{B}$ using the {\it replica trick}.
In the following we assume that the random variables $K_{ij}$ and $\overline{\mu}_i$
are Gaussian distributed with average $\overline{K}$ and $h$ respectively
and standard deviation $\Delta K$ and $\Delta h$ respectively.
We form $n$ identical copies of the system (the {\it replicas}) and 
the average is calculated for an integer $n$ and a finite number of spins $N$. 
Then, using the well-known formula $\ln x=\lim_{n\rightarrow0}(x^n-1)/n$, 
$\overline{\ln Z}$ is obtained  from the analytic continuation of $\overline{Z^n}$
for $n \rightarrow 0$. Finally, we take the thermodynamic limit $N \rightarrow \infty$. 
Explicitly, $\overline{Z^n}$ is given by:
\begin{equation}
\overline{Z^n}=
\sum_{s_i^\alpha=\pm 1} \exp\left[-\overline{H[s_i^\alpha,n]}\right]
\label{rsb9}
\end{equation}
where $\overline{H[s_i^\alpha,n]}$ is the sum of $n$ independent and identical 
spin Hamiltonians, averaged over the disorder, with Greek indices now numbering the $n$ 
replicas. Computing the average over disorder leads
to coupling between spin-spin-interactions of different replicas.

After some analytics that are detailed in \cite{ahufinger2005}, one finally gets
\begin{eqnarray}
\frac{\overline{f}_\textrm{SK}}{k_\textrm{B} T} & = &
\frac{(K/k_\textrm{B} T)^2}{4} (1-q)^2
-\frac{(\Delta h / k_\textrm{B}T)^2}{2} \label{sk10} \\
& & +\int \textrm{d}z\ \frac{e^{-\frac{z^2}{2}}}{\sqrt{2\pi}} 
\ln\left[2\cosh\frac{\sqrt{K^2q+\Delta h^2} - \overline{h}}{k_\textrm{B}T}\right] \nonumber
\end{eqnarray}
and
\begin{eqnarray}
q & = & \int \textrm{d}z\ \frac{e^{-\frac{z^2}{2}}}{\sqrt{2\pi}}
\tanh^2
\left( \frac{\sqrt{K^2q+\Delta h^2}-\overline{h}}{k_\textrm{B}T} \label{sk11} \right) \\
m & = & \int
\textrm{d}z\  \frac{e^{-\frac{z^2}{2}}}{\sqrt{2\pi}}
\tanh \left( 
\frac{\sqrt{K^2q+\Delta h^2}
-\overline{h}}{k_\textrm{B}T} \right) ~.
\label{sk12}
\end{eqnarray}
These are characteristic values of spinglasses that may be measured in experimental realizations
of the proposed systems.

Finally, a study of stability of the replica method as detailed in \cite{ahufinger2005} shows
that the magnetization constraint specific to our model would not change the occurrence of 
{replica symmetry breaking}, provided the M\'ezard-Parisi approach is valid in
finite range spin glass models.

\section{Conclusion}
\label{sec:conclusion}

In this paper, we have reviewed our recent theoretical works on Fermi-Bose mixtures in disordered
optical lattices. In the strongly correlated regime and under constraints that we have
discussed in detail, the physics of the mixture can be mapped into that of single-species
Fermi composites. This is governed by a Fermi-Hubbard like Hamiltonian with parameters
that can be controlled with accuracy in state-of-the-art experiments on ultracold atoms.

We have shown that the presence of disorder (created by random on-site energies) introduces
further control possibilities and induces an extraordinary rich quantum phase diagram. 
For the sake of conciseness, we have restricted our discussion to a particular regime
that proves to be a case study (more details may be found in Ref.~\cite{ahufinger2005}). 
For weak disorder, we have discussed the phase diagram which corresponds to the formation of Fermi glass, Domain insulator, dirty superfluids and metallic phases.
Numerical calculations support our discussion.

For larger amplitudes of disorder, we have shown that the Hamiltonian reduces to that of
a spinglass, {\it i.e.} a spin system with random exchange terms. In our system,
the fictitious spins are coded by the presence or the absence of particles in each lattice
site. 
The physics of spinglasses is a challenging problem in statistical physics which is 
still unsolved. In particular, two theories are competing: 
the {\it droplet picture} and 
the {\it M\'ezard-Parisi picture}. 
These two theories lead to different predictions and even the nature of ordering in
spinglasses is not known. 
On the one hand, the M\'ezard-Parisi picture, which is assumed to be valid
for long-range spin exchange, predicts the existence of a huge number
of quantum states with very similar energies that all contribute to the low-temperature
physics of spinglasses. 
On the other hand, the droplet picture, which is believed to be valid for short spin exchange, 
assumes the existence of only two pure states connected by spin flip symmetry and excitations 
are domains of constant magnetization with fractal boundaries. As possible experimental
realizations of spinglass systems with controllable parameters, mixtures of ultracold 
fermions and bosons may serve as {\it quantum simulators} to solve the controversy and
shed new light on this extraordinary rich physics.


\section*{Acknowledgments}
This work was supported by the Deutsche Forschungsgemeinschaft (SFB 407, SPP1116 436POL),
the RTN Cold Quantum Gases, ESF PESC QUDEDIS, the Alexander von Humboldt Foundation and 
the Ministerio de Ciencia y Tecnologia (BFM-2002-02588). 
J.Z. from the Polish Government Research Funds under contract PBZ-MIN-008/P03/2003.


\begin{thebibliography}{10}
\bibitem{note:alpha}
Notice that, as discussed in section~\ref{sec:composites}, 
we restrict the analysis to the case $\mu_i^{b}<\alpha<1+\mu_i^{b}$.
\bibitem{note:spins}
Notice that the spins commute in different sites $[s_i,s_j]=0$.
\end{thebibliography}
\end{document}